\documentclass[24pt]{article}
\usepackage{amssymb}
\usepackage{epsfig,caption,graphicx,eepic,epic}
\usepackage{pstricks}
\usepackage{amssymb,amsmath}
\usepackage{multido}
\usepackage{array}
\begin{document}
\def\be{\begin{equation}}
\def\ee{\end{equation}}
\def\ba{\begin{eqnarray}} 
\def\ea{\end{eqnarray}}
\def\nn{\nonumber}
\newcommand{\bbf}{\mathbf}   
\newcommand{\rrm}{\mathrm}
\title{\bf Different types of open quantum systems evolving in a Markovian regime}
\author{Tarek Khalil$^{a}$
\footnote{E-mail address: khaliltarek@hotmail.com}\\ 
and\\
Jean Richert$^{b}$
\footnote{E-mail address: j.mc.richert@gmail.com}\\ 
$^{a}$ Department of Physics, Faculty of Sciences(V),\\
Lebanese University, Nabatieh, Lebanon\\ 
$^{b}$ Institut de Physique, Universit\'e de Strasbourg,\\
3, rue de l'Universit\'e, 67084 Strasbourg Cedex,\\      
France} 
 
\date{\today}
\maketitle 
\begin{abstract}
The interaction between an open quantum system and its environment induces generally memory effects generated by the fact that the response of the system to the environment is not instantaneous. Different physical reasons can be at the origin of an absence of time retardation. We present here a study of systems for which instantaneouness is realized although they do not necessarily follow established Markovian criteria.
\end{abstract}
\maketitle  

\section{Introduction}

The interaction between an open system and its environment generates a response of the system which is due to the coupling between the two parts.  The correlations present in the environment generally induce an action on the system memory on a time scale which may be finite or not~\cite{ad,fle1}. They induce so called memory effects. The process is said to be non-Markovian. A Markovian process is  characterized by a succession of short time actions of the environment on the system which are independent from each other. This can be undertood  as an idealization of a realistic process which corresponds to  the existence of finite time correlations between the system and its environment and the signature of an effective complete loss of backflow from the system to the environment.

A rigorous derivation of the two-time memory kernel in a master equation description of the evolution of a system in interaction with an external system is a delicate task which, among other difficulties, involves the problem of time hierarchies~\cite{sp}. The explicit structure of such a kernel is different for each considered system and often the expression of the master equation which governs the evolution of the system is approximated by a phenomenological kernel of Markovian nature~\cite{gk,gl,lzb,sr}. There exist tests relying on positivity properties of the master equation which governs the evolution of the open system and  allow to distinguish between Markovian and non-Markovian behaviour ~\cite{hal}.

More recently two aspects of the memory problem have been re-examined in different contributions. The first one concerns the effective derivation of non-Markovian transport equations~\cite{fle1,hb,vb,ar,fb,sl,ss,mw,bv}, the second one the characterization of the deviation of a process from a Markovian behaviour~\cite{blp,afp,pcz,hpb,ck} by means of a measure of the strength of memory effects.

In the present work we show that a Markovian behaviour of open quantum systems may have different origins which do not necessarily rely
on the time hierarchy imposed by a Markovian regime but nevertheless verify time instantaneousness.

In section 2 we recall the well known Markovian conditions. In section 3.1  we show on an example how a system evolving in a non-Markovian regime can crossover to a Markovian regime as a consequence of the spectral properties of its environment. In section 3.2 the divisibility property characteristic of Markovian processes is applied to the density operator of the system and its environment. The property is shown to lead to a spectral structure of the environment which works as a sufficient condition leading to a Markovian time evolution of the system. In section 4 we impose the properties of the master equation which governs the evolution of a Markovian system to show that we retrieve the solution found in section 3.2. and a further type of interactions between the system and its environment which leads to a Markovian solution. In section 5 we recall and comment the different results.







\section{General conditions for a Markovian evolution: characteristic times and strength of the interaction between the system and its environment}
 
The time evolution of the density operator of an open Markovian quantum system leads to a Lindblad type of master equation ~\cite{gl} under specific conditions which are related to specific orders of magnitude of the characteristic times of evolution of both the systems $S$ and its environment $E$. These times are respectively $\tau_{s}$ and $\tau_{c}$. The time $\tau_{s}$ governs the  evolution of the system $S$ and $\tau_{c}$ the time over which the temporal correlations of the observables of $E$ which enter the coupling between $S$ and $E$ survive. In the Markovian regime which is characterized by the divisibility constraint these times must obey the condition $\tau_{c}\ll \tau_{s}, \tau_{c} \sim 1/\Delta_{E}$ where $\Delta_{E}$ is the extension of the energy spectrum in $E$. Then, in this regime

\ba
\hat\rho_{E}(t)=\bar\rho_{E}+O (\tau_{c}/\tau_{s})
\label{eq1} 
\ea 
where $\bar\rho_{E}$ corresponds to the density operator of a stationary system and the second term on the r.h.s. in Eq.(1) must be a small correction. 

The origin of the characteristic times $\tau_{c}$ is related to the energy extension  of the spectrum of the environment
$\Delta_{E}\sim1/\tau_{c}$ which is the time over which the time correlations of the part of the interaction which corresponds to the environment survive. The time $\tau_{s}$ is the typical interval of time over which the system itself evolves.

A Markovian evolution corresponds to the case where these times correspnd to two very different scales $\tau_{c}\ll\tau_{s}$, more precisely $\tau_{s}\sim1/(|V|^{2}\tau_{c})$ where $|V|$ is the strength of the interaction between the two systems. In order to generate a Markovian behaviour $|V|$ should be weak.

Under these conditions the evolution of the system $S$ depends on a unique time which is characteristic of a Markovian time evolution. It raises the question whether this uniqueness due to the  absence of memory effects can also be obtained under different physical conditions which are cover or not the aforementioned conditions. 


\section{Spectral  properties of the environment}
 
We consider two cases which show how the structure of the environment space can induce a Markovian behaviour.
 
\subsection{The spectrum of the environment extends over an infinite energy interval}
 
We investigate here the transition from a non-Markovian to a Markovian open system by means of a model which shows the role played by the spectral properties of the environment. In order to do this we rely on a model which has been developed recently ~\cite{wei,heng,wei1,heng1}.\\    

{\bf The model}\\                   

The system $S$ is a set of particles with energies $[e_{i}, i=1,N]$ coupled to an environment $E$ of non interacting bosons or fermions through an interaction $H_{SE}$ characterized by a spectral density $J_{\alpha i j}$ where the index ${\alpha}$ designates the particles in the environment. Working out the master equation which governs the density operator in $S$ space one obtains the Green's functions $G_{ij}^{(1)}(t,t_{0})=\langle [a_{i}(t),a_{j}^{+}(t_{0})]\rangle$ in terms of the commutators or anticommutators of the time dependent creation and annihilation operators $a_{i}^{+}(t)$ and $a_{i}(t)$ in $S$ space and $G_{ij}^{(2)}(\tau,t)=\langle a_{j}^{+}(\tau),a_{i}(t)\rangle$. The propagation operators ${\bf G^{(1)}}(t,t_{0})$ and ${\bf G^{(2)}}(\tau,t)$ are respectively related to the retarded Green's function (${\bf G^{(1)}}(t,t_{0})=i{\bf G^{(r)}}(t,t_{0})$) and the lesser Green's function
(${\bf G^{(2)}}(\tau,t)=-i{\bf G^{(<)}}(\tau,t)$).\\

{\bf Evolution of the Green's functions}\\

The $N*N$ Green's functions ${\bf G^{(1)}}(t,t_{0})$ and ${\bf G^{(2)}}(\tau,t)$ obey the following equations

\ba
\frac{d}{d\tau} {\bf G^{(1)}}(\tau,t_{0})+i {\bf e_{s}}{\bf G^{(1)}}(\tau,t_{0})+\int_{t_{0}}^{\tau}d\tau{'} {\bf v}(\tau,\tau^{'}){\bf G^{(1)}}(\tau^{'},t_{0})=0
\label{eq2} 
\ea
and

\ba
\frac{d}{d\tau} {\bf G^{(2)}}(\tau,t)+i {\bf e_{s}}{\bf G^{(2)}}(\tau,t)+\int_{t_{0}}^{\tau}d\tau^{'} {\bf v}(\tau,\tau^{'}){\bf G^{(2)}}(\tau^{'},t)=\int_{t_{0}}^{\tau}d\tau^{'} {\bf v}(\tau,\tau^{'}){\bf (G^{(1)})^{+}}(\tau^{'},t)
\label{eq3}
\ea
where ${\bf e}_{s}$ is the diagonal $N*N$ eigenenergy matrix of the states in $S$ and the $N*N$ propagators ${\bf v}(\tau,\tau^{'})$ are given by 

\ba
{\bf v}(\tau,\tau^{'})=\sum_{\alpha}\int \frac{d\omega}{2\pi}{\bf J}_{\alpha}(\omega)\exp(-i\omega(\tau-\tau^{'}))
\label{eq4} 
\ea
\\

{\bf Instantaneous and time-delayed action of the environment}\\

In the sequel we fix ${\bf J}={\bf J_{0}}=J_{0}\bf 1$ to be diagonal and constant over the whole energy range $(-\infty, +\infty)$. Then  the propagator ${\bf v}(\tau,\tau^{'})= {\bf J_{0}}\delta(\tau-\tau^{'})$  as well as the Green's function matrix in $S$ space are diagonal and the solutions of (2) and (3) read 

\ba
{\bf G^{(1)}}(\tau,t_{0})=\exp{-i({\bf e_{s}}-i{\bf J_{0}})(\tau-t_{0})}
\label{eq5} 
\ea
with ${\bf G^{(1)}}(t_{0},t_{0})=1$ and 

\ba
{\bf G^{(2)}}(\tau,t)= {\bf J_{0}}(\tau-t_{0}) \exp{-i({\bf e_{s}}-i{\bf J_{0}})(\tau-t_{0})}
\label{eq6} 
\ea
with ${\bf G^{(2)}}(t_{0},t)=0$. 

This result shows that $|{\bf G^{(1)}}(\tau,t_{0})|$ decays exponentially as a function of ${\bf J_{0}}$ which is characteristic for Markovian processes. Non-Markovian effects appear in the solution of Eq.(2) if the spectral density function ${\bf J}(\omega)$ shows a 
non constant contribution over the frequency range of $\omega$. As an example consider 

\ba
{\bf J(\omega)}={\bf J_{0}}+\frac{{\bf J_{1}}\Gamma^{2}}{(\omega-E_{0})^{2}+\Gamma^{2}}\Theta(\Omega-|\omega-E_{0}|)
\label{eq7} 
\ea
where $\Omega$ is a band cut-off. If ${\bf J_{1}}=J_{1}\bf 1$ is chosen to be diagonal and $\Omega\rightarrow \infty$ the Green's function reads~\cite{wei1}

\ba
{\bf G^{(1)}}(t-t_{0})= \frac{\exp{-i/2[({\bf e_{s}}-i{\bf J_{0}}+E_{0}-i\Gamma)(t-t_{0})]}}
{2[({\bf e_{s}}-i{\bf J_{0}}-E_{0}+i\Gamma)^{2}+4 {\bf J_{1}}\Gamma]^{1/2}} [\Psi_{-}(t-t_{0})-\Psi_{+}(t-t_{0})]
\label{eq8} 
\ea
where 

\ba
\Psi_{-}(t-t_{0})=[[({\bf e_{s}}-i{\bf J_{0}}-E_{0})+(({\bf e_{s}}-i{\bf J_{0}}-E_{0}+i\Gamma)^{2}+4 {\bf J_{1}}\Gamma)^{1/2}] 
+i\Gamma]
\notag\\
\exp{-i/2[({\bf e_{s}}-i{\bf J_{0}}-E_{0}+i\Gamma)^{2}+4 {\bf J_{1}}\Gamma]^{1/2}(t-t_{0})}
\label{eq9}   
\ea
and

\ba
\Psi_{+}(t-t_{0})= [[({\bf e_{s}}-i{\bf J_{0}}-E_{0})-(({\bf e_{s}}-i{\bf J_{0}}-E_{0}+i\Gamma)^{2}+4 {\bf J_{1}}\Gamma)^{1/2}]
+i\Gamma]
\notag\\
\exp{+i/2[({\bf e_{s}}-i{\bf J_{0}}-E_{0}+i\Gamma)^{2}+4 {\bf J_{1}}\Gamma]^{1/2}(t-t_{0})}
\label{eq10} 
\ea
If the expression of $|{\bf G^{(1)}}(t-t_{0})|$ is worked out one observes oscillatory contributions which add to an exponential decay. This is the signature of the existence of memory effects due to a non-Markovian behaviour. In fact different regimes may set in depending on the strength of the delayed response. In the limit where ${\bf J}_{1}$ tends to zero one retrieves the Markovian limit derived 
above. If the width $\Gamma$ of the resonance located at $E_{0}$ goes to zero, 
${\bf G^{(1)}}(t-t_{0}) \rightarrow \exp{-i/2({\bf e_{s}}-i{\bf J_{0}})(t-t_{0})}$. 
\\ 
 
{\bf Analysis of the behaviour of the Green's function}\\

It is of interest to analyze the dependence of ${\bf G^{(1)}}(t-t_{0})$ on the spectral strength ${\bf J_{1}}$. In order to do this analysis the expression (8) is rewritten in the form

\ba
{\bf G^{(1)}(J_{1})}={\bf A_{1}(J_{1})}\exp\{\bf \Phi_{1}( J_{1})\}-{\bf A_{2}(J_{1})}\exp\{\bf \Phi_{2}( J_{1})\}
\label{eq11} 
\ea
 
where the amplitudes read

\ba
{\bf A_{1}({\bf J_{1}})}=\frac{1}{2}[1+\frac{({\bf E_{-}}-i{\bf V})}{[({\bf E_{-}}-i{\bf V})^{2}+4\Gamma {\bf J_{1}}]^{1/2}}]
\label{eq12} 
\ea
and 

\ba
{\bf A_{2}(J_{1})}=1 - {\bf A_{1}(J_{1})}
\label{eq13} 
\ea
with ${\bf E_{-}}={\bf e_{s}}-E_{0}$, ${\bf V}={\bf J_{0}}-\Gamma$  and the phases come out as

\ba
{\bf \Phi_{1}({\bf J_{1}})}=-\frac{i({\bf E_{+}}-i{\bf W})+i[({\bf E_{-}}-i{\bf V})^{2}+4\Gamma {\bf J_{1}}]^{1/2}}{2}(t-t_{0})
\label{eq14} 
\ea
and

\ba
{\bf \Phi_{2}({\bf J_{1}})}=-\frac{i({\bf E_{+}}-i{\bf W})-i[({\bf E_{-}}-i{\bf V})^{2}+4\Gamma {\bf J_{1}}]^{1/2}}{2}(t-t_{0})
\label{eq15} 
\ea
with ${\bf E_{+}}={\bf e_{s}}+E_{0}$ and ${\bf W}={\bf J_{0}}+\Gamma$.

Define 

\ba
{\bf C}=[({\bf E_{-}}^{2}-V^{2}+4{\bf J_{1}}\Gamma)^{2}+4{\bf E_{-}}^{2}V^{2}]^{1/4}
\label{eq16} 
\ea
and 
 
\ba
\theta=-\arctan[\frac{2E_{-}V}{{\bf E_{-}}^{2}-V^{2}+4{\bf J_{1}}\Gamma}]
\label{eq17} 
\ea
Then

\ba
{\bf \Phi_{1}({\bf J_{1}})}=-\frac{i[({\bf E_{+}}+{\bf C}\cos\theta/2)-i({\bf W}-{\bf C})\sin\theta/2]}{2}(t-t_{0})
\label{eq18} 
\ea
and 
 
\ba
{\bf \Phi_{2}({\bf J_{1}})}=-\frac{i[({\bf E_{+}}-{\bf C}\cos\theta/2)-i({\bf W}+{\bf C})\sin\theta/2]}{2}(t-t_{0})
\label{eq19} 
\ea 
\\
  
{\bf Conclusions}\\
  
\begin{itemize}

\item When ${\bf J_{1}}$ runs from $0$ to $\infty$ the absolute value of the amplitude ${|\bf A_{1}({\bf J_{1}})|}$ decreases from $1$ to $1/2$ and ${|\bf A_{2}(J_{1})|}$ increases from $0$ to $1/2$, both amplitudes showing oscillations over the interval $[0,\infty)$ for any fixed time $t$.

\item The phases ${\bf \Phi_{1}({\bf J_{1}})}$ and ${\bf \Phi_{2}({\bf J_{1}})}$ contain a real and an imaginary part which correspond to an imaginary and a real contribution. 

For both phases the real part leads to an exponential decay if ${\bf W}>{\bf C}$ which is realized for specific values of ${\bf J_{1}}$ depending on the signs and the strengths of $({\bf E_{-}}, V)$ and the strength of $\Gamma$. The imaginary parts oscillate around $E_{+}$ as a function of $\theta/2$ with a frequency which depends on ${\bf J_{1}}$.

For ${\bf J_{1}}=0$ and $\Gamma=0$ these phases read

\begin{center}
${\bf \Phi_{1}}= -i({\bf e_{s}}-i{\bf J_{0}})(t-t_{0})$ and ${\bf \Phi_{2}}= -iE_{0}(t-t_{0})$
\end{center}
 
Since ${\bf A_{2}(J_{1})}=0$ for ${\bf J_{1}}=0$ the behaviour of ${\bf G^{(1)}(J_{1})}$ is independent of ${\bf \Phi_{2}}$ for this value of ${\bf J_{1}}$.

\end{itemize}

\subsection{The spectrum of the environment is restricted to a unique state} 
 
We introduce a general explicit form of the density operator in $S+E$ space and look for the expression of the density operator of the system $S$ which obeys the time divisibility condition.\\

{\bf Divisibility of the time evolution operator}\\
 
Consider a system {\it S} characterized by a density operator  $\hat\rho_{S}(t)$ which evolves in time from 
$t_{0}$ to $t$ under the action of the evolution operator $T(t,t_{0})$

\ba
\hat\rho_{S}(t)=T(t,t_{0})\hat\rho_{S}(t_{0})
\label{eq20}  
\ea

A criterion which may characterize a Markovian behaviour of the system is the divisibility property which reads~\cite{ar,ha,bv} 

\ba
\hat\rho_{S}(t)=T(t,\tau)T(\tau,t_{0})\hat\rho_{S}(t_{0})
\label{eq21}
\ea
for $\tau$ in the interval $[t_{0},t]$, i.e. the operator $T$ possesses a divisibility property. If Eq.(21) is verified the expression of $T$ which acts between $t_{0}$ and $t$ can be split into two completely positive and trace conserving maps from $t_{0}$ to $\tau$ and from $\tau$
to $t$.

At the initial time $t_{0}$  the system $S$ is supposed to be decoupled from the environment and characterized by the density 
operator

\ba
\hat\rho_{S}(t_{0})= \sum_{i_{1},i_{2}}c_{i_{1}}c_{i_{2}}^{*}|i_{1}\rangle \langle i_{2}|
\label{eq22}
\ea
and its environment $E$ by

\ba
\hat\rho_{E}(t_{0})= \sum_{\alpha_{1},\alpha_{2}}d_{\alpha_{1},\alpha_{2}}|\alpha_{1}\rangle \langle \alpha_{2}|
\label{eq23}
\ea
where $|i_{1}\rangle, |i_{2}\rangle $ and $|\alpha_{1}\rangle, |\alpha_{2}\rangle$ are orthogonal states in  
$S$ and $E$ spaces respectively, $c_{i_{1}},c_{i_{2}}$ normalized amplitudes and $d_{\alpha_{1},\alpha_{2}}$ weights such that $\hat\rho_{E}^{2}(t_{0})=\hat\rho_{E}(t_{0})$.

In the following conditions which allow the realization of the equality given by Eq.(21) are derived.\\ 
 
{\bf Density operator for the open system S in the Liouvillian formalism}\\

The complete system $S+E$ is governed by the Hamiltonian
$\hat H= \hat H_{S}+\hat H_{E}+\hat H_{SE}$ where $\hat H_{SE}$ is an arbitrary interaction which couples $S$ to
$E$.
 At the initial time $t_{0}$ the wave function of $S$, $|\psi_{S}(t_{0})\rangle=\sum_{i_{1}}c_{i_{1}}|i_{1}\rangle$, is a 
superposition of orthonormal eigenstates $|i_{1}\rangle$ of $\hat H_{S}$ and the environment is described by the 
density operator $\hat\rho_{E}(t_{0})$ defined above.

If the density operator of the whole system $S+E$ is $\hat\rho(t)$ the reduced density operator in $S$ space 
$\hat\rho_{S}(t)=Tr_{E}[\hat\rho(t)]$ can be written as~\cite{vlb}

\ba
\hat\rho_{S}(t)=\sum_{i_{1},i_{2}}c_{i_{1}}c_{i_{2}}^{*}\hat\Phi_{i_{1},i_{2}}(t,t_{0})
\label{eq24}
\ea

with

\ba
\hat\Phi_{i_{1},i_{2}}(t,t_{0})=\sum_{j_{1},j_{2}}C_{(i_{1},i_{2}),(j_{1},j_{2})}(t,t_{0})|j_{1}\rangle_{S}\langle j_{2}|
\label{eq25}
\ea

where the super matrix $C$ reads

\ba
C_{(i_{1},i_{2}),(j_{1},j_{2})}(t,t_{0})=\sum_{\alpha_{1},\alpha_{2},\gamma}d_{\alpha_{1},\alpha_{2}}
U_{(i_{1}j_{1}),(\alpha_{1}\gamma)}(t,t_{0}) U_{(i_{2}j_{2}),(\alpha_{2}\gamma)}^{*}(t,t_{0})
\label{eq26}
\ea

and

\ba
U_{(i_{1}j_{1}),(\alpha_{1}\gamma)}(t,t_{0})=\langle j_{1}\gamma|U(t,t_{0})|i_{1} \alpha_{1}\rangle
\notag\\
U^{*}_{(i_{2}j_{2}),(\alpha_{2}\gamma)}(t,t_{0})=\langle i_{2} \alpha_{2}|U^{*}(t,t_{0})|j_{2} \gamma \rangle
\label{eq27}
\ea
The evolution operator reads $U(t,t_{0})=e^{-i\hat H(t-t_{0})}$ and the super matrix $C$ obeys the condition 
$\lim_{t\rightarrow t_{0}}C_{(i_{1},i_{2}),(j_{1},j_{2})}(t,t_{0})=\delta_{i_{1},i_{2}}
\delta_{j_{1},j_{2}}$.

In the present formulation the system is described in terms of pure states. The results which will be derived below 
remain valid if the initial density operator at the initial time is composed of mixed states 
$\hat\rho_{S}(t_{0})=\sum_{i_{1},i_{2}}c_{i_{1}i_{2}}|i_{1}\rangle _{S}\langle i_{2}|$.\\

{\bf Imposing the divisibility constraint}\\

The aim is now to find conditions under which the general expression of $\hat\rho_{S}(t)$ obeys the
divisibility constraint imposed by Eq.(21) at any time $t>t_{0}$. 

For this to be realized the following relation must be verified by the super matrix $C$

\ba
C_{(i_{1},i_{2}),(k_{1},k_{2})}(t,t_{0})= \sum_{j_{1},j_{2}}C_{(i_{1},i_{2}),(j_{1},j_{2})}(t_{s},t_{0})
C_{(j_{1},j_{2}),(k_{1},k_{2})}(t,t_{s})
\label{eq28}
\ea
Writing out explicitly the r.h.s. and l.h.s. of eq.(7) for fixed values of $i_{1}$ and $i_{2}$ the 
relation (28) takes the explicit form

\ba
\sum_{\alpha_{1},\alpha_{2},\gamma}d_{\alpha_{1},\alpha_{2}}U_{(i_{1}k_{1}),(\alpha_{1}\gamma)}(t-t_{0})
U^{*}_{(i_{2}k_{2}),(\alpha_{2}\gamma)}(t-t_{0})=
\sum_{j_{1},j_{2}}\sum_{\alpha_{1},\alpha_{2},\beta_{1},\beta_{2}}d_{\alpha_{1},\alpha_{2}} 
d_{\beta_{1},\beta_{2}}
\notag \\
\sum_{\gamma,\delta}U_{(j_{1}k_{1}),(\beta_{1}\delta)}(t-t_{s})U_{(i_{1}j_{1}),(\alpha_{1}\gamma)}(t_{s}-t_{0})
U^{*}_{(j_{2}k_{2}),(\beta_{2}\delta)}(t-t_{s})U^{*}_{(i_{2}j_{2}),(\alpha_{2}\gamma)}(t_{s}-t_{0}) 
\label{eq29}
\ea

In order to find a solution to this equality and without loss of generality we consider the case where the 
density matrix in $E$ space is diagonal. Then the equality reads 
 
\ba
\sum_{\alpha, \gamma}d_{\alpha,\alpha}U_{(i_{1}k_{1}),(\alpha \gamma)}(t-t_{0})
U^{*}_{(i_{2}k_{2}),(\alpha \gamma)}(t-t_{0})=
\sum_{j_{1},j_{2}}\sum_{\alpha,\beta}d_{\alpha,\alpha} d_{\beta,\beta}
\notag \\
\sum_{\gamma,\delta}U_{(j_{1}k_{1}),(\beta \delta)}(t-t_{s})U_{(i_{1}j_{1}),(\alpha \gamma)}(t_{s}-t_{0})
U^{*}_{(j_{2}k_{2}),(\beta \delta)}(t-t_{s})U^{*}_{(i_{2}j_{2}),(\alpha \gamma)}(t_{s}-t_{0}) 
\label{eq30}
\ea

A sufficient condition to realize the equality is obtained if $d_{\beta,\beta}=d_{\alpha,\alpha}$ and consequently if
the weights $d$ on both sides are to be the same one ends up with $d_{\alpha,\alpha}=1$. This last condition imposes a 
unique state in $E$ space, say $|\eta \rangle$. In this case $d_{\eta,\eta}=1$ and eq.(11) reduces to  
 
\ba
U_{(i_{1}k_{1}),(\eta \eta)}(t-t_{0})U^{*}_{(i_{2}k_{2}),(\eta \eta)}(t-t_{0})=
\sum_{j_{1}} U_{(i_{1}j_{1}),(\eta \eta)}(t_{s}-t_{0})U_{(j_{1}k_{1}),(\eta \eta)}(t-t_{s})
\notag\\
\sum_{j_{2}} U^{*}_{(j_{2}k_{2}),(\eta \eta)}(t-t_{s})U^{*}_{(i_{2}j_{2}),(\eta \eta)}(t_{s}-t_{0}) 
\label{eq31}
\ea
which proves the equality.\\

A further property imposed by the divisibility constraint can be observed if the evolution operator $e^{-i\hat H (t-t_{0})}$ 
is developed in a factorized product of exponential terms~\cite{za} 

\ba
e^{-i(\hat H_{0}(t-t_{0})+\hat H_{SE}(t-t_{0}))}=e^{-i\hat H_{0}(t-t_{0})}e^{-i\hat H_{SE}(t-t_{0})}
\hat \Omega[(t-t_{0}), \hat H_{0},\hat H_{SE}]
\label{eq32}
\ea
where $\hat H_{0}=\hat H_{S}+\hat H_{E}$ and $\hat \Omega[(t-t_{0}),\hat H_{0},\hat H_{SE}]$ is generally an infinite product of exponentiated commutators. The eigenstates of the system space $\cal H_{S}$ are  chosen to be the basis states with eigenvalues $[\epsilon_{i}]$.

The equality of the r.h.s.and l.h.s. of Eq.(28) can be realized in  the special case where  $\hat \Omega=1$ or 
$\hat \Omega=e^{ic \hat 1}$ where $c$ is a real number. Then for fixed $(i_{1},i_{2}),(k_{1},k_{2})$ the equality (31) reads

\ba 
\lefteqn{\exp[-i(\epsilon_{k_{1}}-\epsilon_{i_{2}})(t-t_{0})
\langle k_{1}\eta|\exp(-i\hat H_{SE}(t-t_{0}))|i_{1}\eta\rangle
\langle i_{2}\eta|\exp(i\hat H_{SE}(t-t_{0}))|k_{2}\eta\rangle =}
\nonumber\\ 
&  & \sum_{j_{1}}
\exp[-i[(\epsilon_{k_{1}}+E_{\eta})(t-t_{s})+(\epsilon_{j_{1}}+E_{\eta})(t_{s}-t_{0})]
\langle j_{1}\eta|\exp(-i\hat H_{SE}(t_{s}-t_{0}))|i_{1}\eta\rangle
\nonumber\\
& & \langle k_{1}\eta|\exp(-i\hat H_{SE}(t-t_{s}))|j_{1}\eta\rangle*
\nonumber\\
& & \sum_{j_{2}}
\exp[+i[(\epsilon_{j_{2}}+E_{\eta})(t-t_{s})+(\epsilon_{i_{2}}+E_{\eta})(t_{s}-t_{0})]
\langle j_{2}\eta|\exp(i\hat H_{SE}(t-t_{s}))|k_{2}\eta\rangle
\nonumber\\
& & \langle i_{2}\eta|\exp(i\hat H_{SE}(t_{s}-t_{0}))|j_{2}\eta\rangle
\label{33}
\ea
One notices that the eigenenergy $E_{\eta}$ is eliminated on both sides of the equality. 

Furthermore the equality between the l.h.s. and the r.h.s. of eq.(14) is now realized if the following constraints are satisfied:

\begin{itemize}

\item  $\epsilon_{j_{1}}=\epsilon_{k_{1}}$ and $|j_{1}\rangle=|k_{1}\rangle$ or
$|j_{1}\rangle \bot |k_{1}\rangle$, i.e. if the states are degenerate in energy.

\item $\epsilon_{j_{2}}=\epsilon_{i_{2}}$and $|j_{2}\rangle=|i_{2}\rangle$ or
$|j_{2}\rangle \bot |i_{2}\rangle$ if the states are degenerate in energy.

\end{itemize}

{\bf Physical implications of the solution}\\

The present analysis reveals the existence of specific systems which obey the Markovian divisibility property even if they do not necessarily follow the criteria developed in section 2.

\begin{itemize}
 
\item The sufficient condition introduced above imposes that at time $t_{0}$  the environment $E$ has to be in a fixed eigenstate $|\eta\rangle$ with probability $1$. The state may  be a ground state or an excited state. It has to evolve in time by staying in this initial state. This property may be realized in practice if $E$ is a thermal environment which stays at a temperature close to $T=0$.

If the truncated factorization development introduced by eq.(32) above works the phases appearing in the factorized r.h.s. before and after an arbitrary time $t_{s}$ are the same in both intervals $[t_{0},t_{s}]$ and $[t_{s},t]$ ($j_{n}=i_{n}$ or $j_{n}=k_{n}$ ). Hence phase interference effects due to different eigenstates of $S$ do not appear for any intermediate time $t_{s}$, the phases generated by the time evolution of the system $S$ stay the same before and after this time. It has been shown ~\cite{blp} that Markov processes tend continuously to reduce  the distinguishability of any two states. Indeed, if two states of $S$ are degenerate at time $t_{0}$ they will stay degenerate, hence their distance ~\cite{nc} will stay equal to zero over any interval of time.

\item  Another interesting point concerns the spectrum of $S$. It comes out from the examination of the introduced 
divisibility constraints that for the same energies $\epsilon_{j_{1}}=\epsilon_{k_{1}}$ and $\epsilon_{j_{2}}=\epsilon_{i_{2}}$ the corresponding states can be either the same or orthogonal to each other. In the last case two orthogonal states are degenerate in energy. This property is also characteristic of the location of a quantum phase transition~\cite{scs,kr}.

This fact which links a rigorous Markovian behaviour to quantum criticality appears here for general Hamiltonians 
$\hat H= \hat H_{S}+\hat H_{E}+\hat H_{SE}$ for which the truncated factorization relation works. It can be confronted with similar findings obtained on hand of models by H. T. Quan et al.~\cite{htq} and more recently by P. Haikka et al.~\cite{ha} which show this remarkable correlation as well as the fact that at the same time the Loschmidt echo is vanishing at this point. The explanation for this characteristic behaviour may be related to the property of systems which do not keep memory of their past and criticality. 
At critical points systems are in an intermediate stage between two phases and belong neither to one phase nor to the other. At such points the memory of the structure of the system coming from a specific phase is lost at any time and memory loss is also the essence of the systems showing Markovian properties.

\end{itemize}

{\bf Entangled initial conditions}\\  

Consider the more general case for which initial correlations at $t_{0}$ are present~\cite{bu}. Then the initial density operator can be written as $\hat \rho (t_{0}) = |\Psi(t_{0})\rangle \langle \Psi(t_{0})|$ with$|\Psi(t_{0})\rangle= \sum_{i,\alpha} a_{i,\alpha}|i,\alpha\rangle$.
Using the same notations as above the component ($k_{1}, k_{2}$) of $\hat \rho^{k_{1} k_{2}}_{S}(t)$ reads 

\ba
\hat\rho_{S}^{k_{1} k_{2}}(t)= \sum_{i_{1},i_{2}}\sum_{\alpha_{1},\alpha_{2},\delta} a_{i_{1},\alpha_{1}} a_{i_{2},\alpha_{2}}^{*}
U_{(i_{1}k_{1}),(\alpha_{1}\delta)}(t,t_{0})U^{*}_{(i_{2}k_{2}),(\alpha_{2}\delta)}(t,t_{0})|k_{1}\rangle \langle k_{2}|
\label{eq34} 
\ea

The divisibility criterion imposes

\ba
\sum_{j_{1},j_{2}}\sum_{\eta}\sum_{\gamma}U_{(j_{1}k_{1}),(\eta\gamma)}(t,t_{s})U^{*}_{(j_{2}k_{2}),(\eta\gamma)}(t,t_{s})
\notag\\
\sum_{\delta}U_{(i_{1}j_{1}),(\alpha_{1}\delta)}(t_{s},t_{0})U^{*}_{(i_{2}j_{2}),(\alpha_{2}\delta)}(t_{s},t_{0})|k_{1}\rangle \langle k_{2}|=
\notag\\
\sum_{\epsilon} U_{(i_{1}k_{1}),(\alpha_{1}\epsilon)}(t,t_{0})U^{*}_{(i_{2}k_{2}),(\alpha_{2}\epsilon)}(t,t_{0})|k_{1}\rangle \langle k_{2}|
\label{eq35}
\ea

A sufficient condition in order to obtain the equality of the two sides in Eq.(35) is realized if the summation of the states are 
such that $|\delta\rangle=|\eta\rangle$. Then the summation over the intermediate states $j_{1},j_{2}$ on the l.h.s. of Eq.(35) 
can only be performed independently if the summation over $E$ space reduces to a unique state which guarantees the possible use of the closure property in order to sum over the intermediate states in $S$ space. Hence the present solution leading to the divisibility 
property does no longer hold when the system $S$ and the environment $E$ are already interacting at the initial time $t_{0}$ 
except if $E$ space contains a unique state, say $|\delta\rangle$. The correlation between the initial state of the system and a non-Markovian behaviour of the time evolution of the system has been demonstrated recently by means of different arguments ~\cite{rmma}.\\

{\bf Entropy properties of the systems}\\

The interaction Hamiltonian $H_{SE}$ generates entanglement between the system $S$ and the environment $E$. On the other hand this coupling is also the source of time retardation (non-Markovian memory) effects in the time behaviour of the system $S$. One may ask how the absence of retardation imposed by the strict divisibility constraint is correlated with the entanglement induced by the coupling between the two systems.\\

When divisibility is strictly verified by means of the sufficient condition found above the matrix elements of $\hat\rho_{S}(t)$ takes the form

\ba
\rho_{S}^{j_{1},j_{2}}(t)=\sum_{i_{1},i_{2}}c_{i_{1}}c_{i_{2}}^{*}
\langle j_{1}\eta|U(t,t_{0})|i_{1} \eta\rangle \langle i_{2} \eta|U^{*}(t,t_{0})|j_{2} \eta\rangle |j_{1}\rangle \langle j_{2}| 
\label{eq36}
\ea
In this case one sees that the entanglement is reduced to the coupling of the system to a one-dimensional environment space. The Hilbert space of the total system $S+E$ reduces in practice to dimension $d+1$ where $d$ is the dimension of $S$.\\ 
 

A test concerning the time evolution of entanglement in an open quantum system which rely on a conjecture of Kitaev have been worked out recently~\cite{aco} which proves the so called "small incremental entangling" (SIE)~\cite{bra}. 

It was shown that in the absence of ancilla states the maximum time evolution of the von Neumann entropy 
$\Sigma_{S}(t)=-Tr \hat\rho_{S}(t)\log \hat\rho_{S}(t)$ verifies 

\ba
\Gamma_{max}=\frac{d\Sigma_{S}(t)}{dt}|_{t=0}\leq c \|H\|\log \delta
\label{eq37}
\ea
where $\delta=min(d_{S}, d_{E})$, the smallest dimension of $S$ and $E$ space, $\|H\|$ is the norm of the Hamiltonian and $c$ a constant of the order of unity.

In the present case $\delta=1$, hence $\Gamma_{max}=0$ which shows that the entropy of the considered here stays constant over time. There is no change in the information content of $S$ in this case.\\

{\bf Characterization of a Markovian regime}

One may now make recall the conditions for a Markovian behaviour presented at the beginning of section 2 in Eq.(1) and confront them with the last two cases:

\begin{itemize}

\item If there is only one state in $E$ $\hat\rho_{E}(t)=\bar\rho_{E}$ since there is only one state present in $E$ space.  Here there is no restriction on the strength of the coupling interaction $|V|$ which can be arbitrarily large.

\item If the energy spectrum in $E$ extends over a very large energy interval $\Delta_{E}$ $\tau_{c}$ tends to zero and the second term in Eq.(1) goes to zero when the spectrum extends to infinity.\\

\end{itemize}






In the next section we rely on the master equation of the system in order to show that the specific structure of the environment leads to a solution which is consistent with the solution found above and leads to a further class of interactions which satisfy a Markovian behaviour.\\ 

\section{Solutions of the master equation: structure of the environment and properties of the coupling interaction}

{\bf Expression of the master equation}\\

The density operator of an open quantum system in a time local regime can be written in the form ~\cite{gk}

\ba
\frac{d}{dt}\hat\rho_{S}(t)=\sum_{n}L_{n}(t)\hat\rho_{S}(t)R_{n}^{+}(t)
\label{eq38}
\ea
where $L_{n}(t)$ and $R_{n}(t)$ are time local operators.

Using the general expression of the density operator $\hat\rho_{S}(t)$ given by Eqs. (24-27) above and taking its time derivative leads to two contributions to the matrix elements of the operator 

\ba
\frac{d}{dt}\rho_{S1}^{j_{1}j_{2}}(t)=(-i)\sum_{i_{1}i_{2}}c_{i_{1}}c^{*}_{i_{2}}\sum_{\alpha_{1},\alpha_{2}}d_{\alpha_{1},\alpha_{2}}
\sum_{\beta \gamma k_{1}}\langle j_{1}\gamma|\hat H|k_{1} \beta \rangle
\notag\\
\langle k_{1}\beta|e^{-i\hat H(t-t_{0})}|k_{1} \beta \rangle \langle i_{2}\alpha_{2}|e^{i\hat H(t-t_{0})}|j_{2} \beta \rangle 
\label{eq39}
\ea
and

\ba
\frac{d}{dt}\rho_{S2}^{j_{1}j_{2}}(t)=(+i)\sum_{i_{1}i_{2}}c_{i_{1}}c^{*}_{i_{2}}\sum_{\alpha_{1},\alpha_{2}}d_{\alpha_{1},\alpha_{2}}
\sum_{\beta \gamma k_{2}}\langle j_{1}\gamma|e^{-i\hat H(t-t_{0})}|i_{1} \alpha \rangle  
\notag\\
\langle i_{2}\alpha_{2}|e^{i\hat H(t-t_{0})}|k_{2} \beta \rangle \langle k_{2}\beta|\hat H|j_{2} \gamma \rangle 
\label{eq40}
\ea
where                  

\ba
\frac{d}{dt}\hat\rho_{S}^{j_{1}j_{2}}(t)=\frac{d}{dt}\rho_{S1}^{j_{1}j_{2}}(t) + \frac{d}{dt}\rho_{S2}^{j_{1}j_{2}}(t)
\label{eq41}
\ea
From the explicit expression of the density operator and Eqs.(39-41) it can be seen that the structure of the master equation given by 
Eq.(38) can only be realized if $|\beta \rangle$ is identical to $|\gamma \rangle$. This constraint has three solutions:

\begin{itemize}

\item There is only one state $|\gamma \rangle$ in $E$ space. This result has already been seen on the expression of the density operator above.

\item The Hamiltonian $\tilde H=\hat H_{E}+\hat H_{SE}$ is diagonal in $E$ space, $[\hat H_{E},\hat H_{SE}]=0$. See details in Appendix A.

\item The density operator $\hat\rho_{S}(0)$ is diagonal in $S$ space with equal amplitudes of the states and the states in $E$ space are equally weighed, $\hat \rho_{E}=\sum_{\alpha}d_{\alpha,\alpha}|\alpha\rangle\langle\alpha|$, $d_{\alpha,\alpha}=1/N$ where $N$ is the number of states in E space. See details in Appendix B.

\end{itemize}

All three conditions are sufficient to insure the structure of the  r.hs. of Eq.(38) and the evolution of the density operator can be written as 

\ba
\frac{d}{dt}\hat\rho_{S1}(t)=\hat O \otimes \hat\rho_{S}(t) \otimes \hat I
\notag\\
\frac{d}{dt}\hat\rho_{S2}(t)= \hat I \otimes \hat\rho_{S}(t) \otimes \hat O^{+}
\label{eq42}
\ea
where $\hat I$ is the identity operator in $S$ space and 

\ba
\hat O = (-i)(\hat D_{S}+ \hat D_{SE} + \hat N_{SE})
\label{eq43}
\ea
The operator $\hat D_{S}$ corresponds to $\hat H_{S}$ in a basis of states in $S$ space in which it is diagonal, $\hat D_{SE}$ is the diagonal part and $\hat N_{SE}$ the non-diagonal part of the matrix which corresponds to $\hat H_{SE}$.\\

The master equation can be written in operator form

\ba
\frac{d}{dt}\hat\rho_{S}(t)=\hat O \hat\rho_{S}(t)+\hat\rho_{S}(t)\hat O^{+}
\label{eq44}
\ea
The trace of $\hat\rho_{S}(t)$ being conserved it follows that $Tr(\hat O\hat\rho_{S}(t)+\hat\rho_{S}(t)\hat O^{+})=0$ and 
consequently $\hat O +\hat O^{+}=0$. Following~\cite{hal} one obtains 

\ba
\frac{d}{dt}\hat\rho_{S}(t)=[\hat O -\hat O^{+},\hat\rho_{S}(t)]
\label{eq45}
\ea 
with $\hat O -\hat O^{+}=\hat H_{S}+\hat H^{nd}_{SE}$ and $\hat H^{nd}_{SE}$ is the non-diagonal part of $\hat H_{SE}$.\\ 
 
{\bf Comments}\\ 
 
The derived master equation leads to  the following properties and comments.

\begin{itemize}
 
\item The property of the density operator with respect to its dependence on the environment is consistent with the property derived through the application of the divisibility constraint in section 3.
 
\item The master equation does not show a second term corresponding to the decoherence contribution which appears in a Lindblad equation. Hence decoherence should not occur in this specific case. 
 
\item In agreement with the present result it can be observed in the Feshbach projection approach of ref.~\cite{ck1} that the presence of a unique state in $E$ space leads to an evolution equation which stays local in time. 

\end{itemize}

\section{Summary and conclusions}   
 
In the present work we recalled the general conditions under which an open quantum system evolves in a Markovian regime governed by 
characteristic time scales of the system and the environment and a weak interaction between the two parts.\\

In section 3 it was asked whether these criteria are necessary in order to characterize open systems which are not affected by memory effects.\\

First the Markovian character of the evolution of an open system was examined in the framework of specific models~\cite{wei, heng,heng1} fixed by the structure of the spectral function which characterizes the coupling between the two systems. In this case the spectrum is  a continuum over an infinite range of frequencies. The width of the environment spectrum is correlated with a typical evolution time in the environment which tends to zero. There is no constraint however on the strength of the interaction between the system and its environment. Deviations from this idealized case introduce non-Markovian corrections and the crossover from a Markovian regime to a non-Markovian regime has been investigated in the framework of the model mentioned above. It would be of particular interest to verify these results in a more general framework.\\

In section 3.2 we used the explicit Liouvillian expression of the density operator in order to test the divisibility constraint. It was found that this property can be realized if the spectrum in the environment space reduces to a unique state, independently of any time scale or strength of the interaction beween the system and its environment. The structure is such that there is no possibility for any excitation in its spectrum which impedes retardation or backflow between the two parts of the total system.\\ 
 
As a by-product we showed that in specific cases the Markovian time evolution of the system can be the signature of the presence of a critical point corresponding to a phase transition of first order or second order in an infinite system. This fact can be understood as the consequence of the loss of memory of the system at such a point.\\
 
In the case where the divisibility property is generated by the presence of a unique state in the environment space the conjectured expression of the entanglement between the system and the environment spaces is such that the entanglement entropy is constant over 
time.\\ 

The time evolution of the open system cannot possess the divisibility property if the spectrum of the environment contains more than one state and the wave function of the system and its environment are entangled from the beginning.\\ 

These results are general and do not depend on a specific model.\\

In section 4 we introduced the master equation which governs the evolution of the density operator. We showed that the equation is of Markovian type if there is a unique state in the environment space. This result is coherent with the preceding result concerning divisibility and in  agreement with recent results relying on a Feshbach projection method in order to derive master equations~\cite{ck1}.\\

Finally we found out that possible symmetries of the interaction may also lead to Markovian processes.\\


{\bf Acnowledgments}
J.R. would like to thank Prof. Janos Polonyi for critical comments and fruitful suggestions during the time of elaboration of the present work.

\section{Appendix A} 

The expressions of $\frac{d}{dt}\rho_{S1}^{j_{1}j_{2}}(t)$ and $\frac{d}{dt}\rho_{S2}^{j_{1}j_{2}}(t)$ given in Eqs. (39) and (40) can be written as
 
\ba
\frac{d}{dt}\rho_{S1}^{j_{1}j_{2}}(t)=(-i)\sum_{k_{1}k_{2}}A^{j_{1}k_{1}}_{\gamma}\rho^{k_{1}k_{2}}_{S\gamma}(t)I^{k_{2}j_{2}}
\ea
where $I$ is the identity operator in S space and 
 
\ba
A^{j_{1}k_{1}}_{\gamma}=\langle j_{1}\gamma|\hat H_{S}|k_{1}\gamma\rangle + \langle j_{1}\gamma|\hat H_{E}+\hat H_{SE}|k_{1}\gamma\rangle
\ea 
and similar expressions for $\frac{d}{dt}\rho_{S2}^{j_{1}j_{2}}(t)$. The matrix elements of $\hat H_{SE}$ in the second term on the r.h.s. of the expression of $A$ are generally non diagonal in E. This is however the case iff $\hat H_{E}$ and $\hat H_{SE}$ commute.
 
\section{Appendix B}  
 
Starting from the expression of the density operator given by Eqs.(24)-(27) we consider the case where $|c_{i}|=1/n$  for all $i$ where $n$ is the number of states in $S$ space and $d_{\alpha_{1},\alpha_{2}}=1/N \delta_{\alpha_{1},\alpha_{2}}$.

In this case the relation which imposes the divisibility constraint reads 
 
\ba
\frac{1}{Nn}\sum_{i\alpha,\gamma}U_{(ik_{1}),(\alpha\gamma)}(t,t_{0})U^{*}_{(ik_{2}),(\alpha\gamma)}(t,t_{0})=
\notag\\ 
\frac{1}{N^{2}n}\sum_{j_{1}j_{2}\beta,\delta}U_{(j_{1}k_{1}),(\beta\delta)}(t,t_{s})U^{*}_{(j_{2}k_{2}),(\beta\delta)}(t,t_{s})
\notag\\
\sum_{i,\alpha,\gamma}U_{(ij_{1}),(\alpha\gamma)}(t_{s},t_{0})U^{*}_{(ij_{2}),(\alpha\gamma)}(t_{s},t_{0})
\label{eq46} 
\ea

The expression in the last line leads to  

\ba
\sum_{i,\alpha,\gamma}U_{(ij_{1}),(\alpha\gamma)}(t_{s},t_{0})U^{*}_{(ij_{2}),(\alpha\gamma)}(t_{s},t_{0})= N\delta_{j_{1},j_{2}} 
\label{eq47}
\ea 
 
and finally the r.h.s. reduces to

\ba 
\sum_{j_{1}j_{2}\beta,\delta}U_{(j_{1}k_{1}),(\beta\delta)}(t,t_{s})U^{*}_{(j_{2}k_{2}),(\beta\delta)}(t,t_{s})=1/N\delta_{k_{1},k_{2}} 
\label{48}
\ea 
 
It is easy to see that working out the l.h.s. of Eq.(30) leads to the same result.

\end{document}